\documentclass{iauRR}

\usepackage{amsmath}
\usepackage{graphicx}
\usepackage{multirow}

\def\ve#1{{\bf #1}}

\def\TCX{{\rm TCX}}
\def\TCL{{\rm TCL}}
\def\TCG{{\rm TCG}}
\def\TCB{{\rm TCB}}
\def\TT{{\rm TT}}
\def\TDB{{\rm TDB}}

\def\TAI{{\rm TAI}}

\def\gaia{{\it Gaia}}

\def\ssr{{Space Science Reviews}}
\def\aap{{Astron.~Astrophys.}}
\def\aj{{Astron.J.}}
\def\prd{{Phys.Rev.D}}

\begin{document}

\journaltitle{}
\jnlPage{1}{7}
\jnlDoiYr{2025}
\doival{}

\lefttitle{Sergei A. Klioner}
\righttitle{Relativistic time scales in the Solar system}

\aopheadtitle{}
\editors{}

\title{Relativistic time scales in the Solar system}

\author{Sergei A. Klioner}
\affiliation{Lohrmann Observatory, Technische Universit\"at Dresden, 01062 Dresden, Germany}

\begin{abstract}
This paper summarizes theoretical definitions of the relativistic
coordinate time scales introduced by the IAU 2000 framework as well as
practical aspects of their use.  It is argued that the IAU framework
already defines relativistic local GCRS-like reference systems and the
corresponding \TCG-like coordinates times for each body of the Solar
system. The interrelations between the coordinate times and the proper
time of an observer are discussed. The arguments put forward that any
scaling of the local coordinate times like \TCL\ for the Moon is
unreasonable. Practical recipes of the transformations between
\TCB\ and the local coordinate time scales (\TCG, \TCL, etc) are then
discussed. Time ephemerides giving the transformation between
\TCB\ and the local coordinate times at the center of mass of the
corresponding body are computed for all major bodies of the Solar system
using INPOP19a. Those time ephemerides represented as a standard set
of Chebyshev polynomials are available online.
 \end{abstract}


\maketitle

\section{Relativistic reference systems of the IAU}

The IAU Resolutions 1991 and 2000 augmented by two Resolutions of 2006
\citep{IAU1991,IAU2000,IAU2006} have laid a solid foundation for the
relativistic framework for modelling of astronomical
observations (see e.g. \citet{2003AJ....126.2687S} for detailed
explanations). The main application of this framework is modelling of
the high-accuracy astronomical observations in the Solar system. Over
last decades many important applications of this framework have been
formulated: the model for VLBI observations
\citep{2010ITN....36....1P,2017JGeod..91..783S},
the relativistic model for Gaia observations
\citep{2003AJ....125.1580K,2004PhRvD..69l4001K}, etc.

In the original IAU 2000 framework only two relativistic reference
systems are explicitly defined: the Barycentric Celestial Reference
System (BCRS) and the Geocentric Celestial Reference System
(GCRS). The BCRS is a global reference system physically adequate to
model physical phenomena occurring in the Solar system as a whole. For
example, the BCRS is used to model the planetary motion in the Solar
system. The GCRS is a physically adequate 'local' reference system
suitable to model phenomena in the vicinity of the Earth (say, within
the radius of the geostationary orbit).

In reality, the very same relativistic framework can and should be
used to define physically adequate local reference systems for each
body of the Solar system. Although the IAU 2000 framework doesn't
mention this, the corresponding original publications of the
relativistic framework \citep[see,
  e.g.,][]{1988CeMec..44...87K,1991PhRvD..43.3273D,2000PhRvD..62b4019K}
always consider $N+1$ reference systems for a system of $N$ bodies:
one global system BCRS and $N$ 'local' GCRS-like reference
systems for each of the $N$ bodies.  The IAU 2024 Resolution II \citep{IAU2024}
explicitly introduces the local relativistic reference system for the Moon
referring to the definitions in the IAU 2000 Resolutions and also 
mentions that the same principles can be
applied to other bodies in the Solar system. Along with the Moon,
Mars and Mercury are especially interesting in this respect in the
coming years and decades (see e.g. \citet{FallettaRodridezKlioner2025}
for an application for Mercury). The body for which the local GCRS-like reference system is
constructed will be called 'central body' below.
The local reference systems like GCRS and LCRS have two important properties:
\begin{itemize}
  \setlength{\leftskip}{10pt}
  \setlength{\itemindent}{30pt}
\item[{\bf A.}] The gravitational field of external bodies is
represented only in the form of a relativistic tidal potential which is at
least of second order in the local spatial coordinates and coincides
with the usual Newtonian tidal potential in the Newtonian limit.
\item[{\bf B.}] The internal gravitational field of the central body (Earth, Moon, etc.) 
coincides with the gravitational field of a corresponding isolated
body provided that the tidal influence of the external matter is
neglected.
\end{itemize}
\noindent
The GCRS, the LCRS and similar local reference systems are the
relativistic (post-Newtonian) generalizations of the corresponding
Newtonian quasi-inertial reference systems that co-move with their central bodies and explicitly
implement the maximal possible effacement of external gravitational field as guaranteed by
the Strong Equivalence Principle valid in General Relativity
\citep[e.g.][]{1991PhRvD..43.3273D,2000PhRvD..62b4019K}.

\begin{figure}[h!]
  \begin{center}
    \includegraphics[scale=.45]{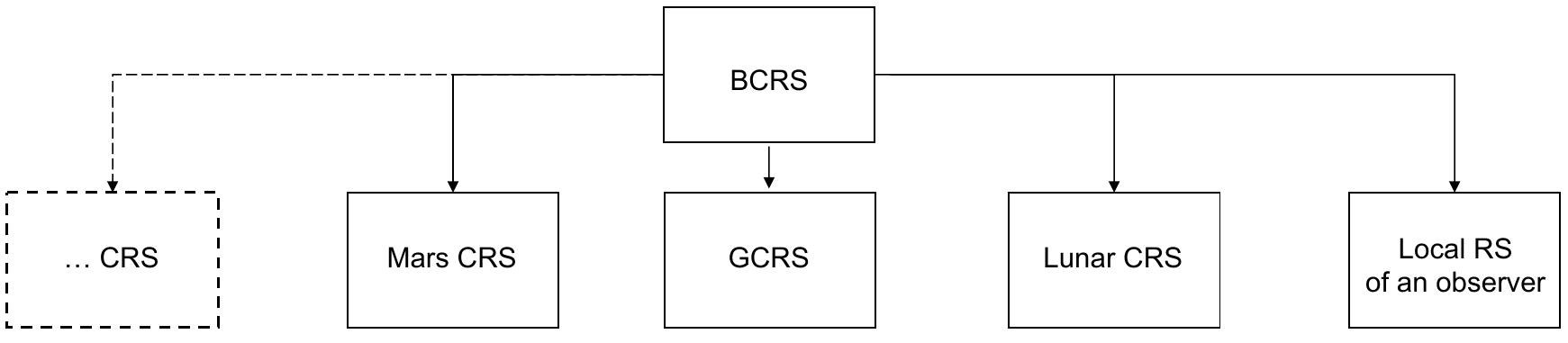}
  \end{center}
  \vspace*{-5mm}
  \caption{The hierarchy of the relativistic reference systems in the IAU 2000 framework.}
  \label{fig:IAU-hierarchy}
\end{figure}

The same principles of constructing the local reference systems can
also be applied to a massless body (a body with negligible mass,
e.g. a spacecraft) to define a physically adequate relativistic
reference system of an observer. Observable quantities like observable
(proper) time and observable directions towards astrometric sources
can be directly computed using that local observer's reference system
\citep[e.g.][]{2004PhRvD..69l4001K}. The hierarchy of relativistic
reference systems to be used for high-accuracy modelling of
observational data can be visualized as shown on
Fig~\ref{fig:IAU-hierarchy}. On the practical side, the relativistic
reference systems can be considered as set of algorithms to model
high-accuracy astronomical observations of various kinds
\citep[see e.g.][]{2011rcms.book.....K,2012asas.book.....V,2019agr..book.....S}.

In the IAU framework, each reference system is given by its metric
tensor $g_{\alpha\beta}$, $\alpha$ and $\beta$ are indices running
from 0 to 3: the time coordinate corresponding to index 0 and the
spatial coordinates correspond to indices 1 to 3.  The metric tensors
allow one to derive the equations of motion (of massive bodies or
light rays) in the corresponding reference systems. Both the metric
tensors and those equations of motion contain parameters like masses
and other multipole coefficients of the gravitational fields as well
as initial positions and velocities of all involved bodies. Those
parameters should be fitted to observations in the process of creation
of the ephemerides of the Solar system or a spacecraft. We don't
discuss this process here and will not further consider the details of
the metric tensors in this publication.

Since the IAU 2000 framework is relativistic it naturally includes both
relativistic time scales and spatial coordinates.  The main topic of
this paper is a discussion of the time scales implied by the IAU
framework. For this discussion the transformations between the BCRS
coordinates $(t=\TCB,x^i)$ and the coordinates $(T=\TCX,X^a)$ of a GCRS-like
reference system 'XCRS' are of interest. Here \TCX\ is generic
name for the coordinate time of the corresponding local reference
system XCRS for any central body $X$. For example for the Earth
$\TCX\equiv\TCG$ and for the Moon $\TCX\equiv\TCL$. These
transformations are explicitly provided by the IAU 2000 framework:
\begin{eqnarray}
\label{T-t-IAU}
T &=& t - {1 \over c^2} \biggl[ A(t) + v_X^i r_X^i\biggr] + {1 \over c^4}
\biggl[B(t) + B^i(t)r_X^i + B^{ij}(t)r_X^i r_X^j + C(t,\ve{x})\biggr ]
+ O(c^{-5}), 
\\
\label{X-x-IAU}
X^a &=& \delta_{ai} \biggl[ r_X^i + {1 \over c^2}
\biggl({1 \over 2} v_X^i  v_X^j
r_X^j + w_{\rm ext}(\ve{x}_X)r_X^i + r_X^i a_X^j r_X^j - {1 \over 2} a_X^i r_X^2
\biggr) \biggr] + O(c^{-4}),
\end{eqnarray}
\noindent
where $c$ is the velocity of light, $r_X^i=x^i-x_X^i$ and the Latin
indices running from 1 to 3 enumerate the spatial coordinates.  The
repeated Latin indices assume the Einstein summation: e.g. $v_X^i
r_X^i\equiv\sum_{i=1}^3v_X^i r_X^i$.  Then one has
\begin{eqnarray}
\label{At}
{d \over dt}A(t) &=& {1 \over 2} v_X^2 + w_{\rm ext}(\ve{x}_X)\,, \\
\label{Bt}
{d \over dt}B(t) &=& -{1 \over 8}
v_X^4 - {3 \over 2} v_X^2 w_{\rm ext}(\ve{x}_X)
+ 4v_X^iw_{\rm ext}^i(\ve{x}_X) + {1 \over 2}w_{\rm ext}^2(\ve{x}_X)\,, \\
B^i(t) &=& -{1 \over 2}v_X^2 v_X^i + 4 w_{\rm ext}^i(\ve{x}_X) -
3v_X^iw_{\rm ext}(\ve{x}_X)\,, \label{eq:Bi}\\
B^{ij}(t) &=&
-v_X^i\delta_{aj}Q^a + 2{\partial \over \partial x^j}w_{\rm ext}^i
(\ve{x}_X) - v_X^i{\partial \over \partial x^j} w_{\rm ext}(\ve{x}_X)
+  {1 \over 2}\delta^{ij}\dot{w}_{\rm ext}(\ve{x}_X)\,, \label{eq:Bij}\\
C(t,{\bf x}) &=& -{1 \over 10}r_X^2(\dot{a}_X^i r_X^i)\,. \label{eq:C}
\end{eqnarray}
\noindent
Here $x_X^i$, $v_X^i$, and $a_X^i$ are the barycentric
position, velocity and acceleration vectors of the central body for which the
reference system $(T,X^a)$ is constructed, the dot
stands for the total derivative with respect to $t$, and
\begin{equation}
Q^a = \delta_{ai} \left[{\partial \over \partial x_i} w_{\rm ext}(\ve{x}_X)
- a_X^i\right]\, .
\end{equation}
\noindent
Here boldface characters signify the components of three-dimensional quantities ${\bf a}=a^i$.
The external potentials, $w_{\rm ext}$ and $w_{\rm ext}^i$, are given by
\begin{equation}
w_{\rm ext} = \sum_{A\not= X}w_A, 
\qquad
w_{\rm ext}^i = \sum_{A\not= X} w_A^i\, ,
\end{equation}
\noindent
where $A$ enumerates solar system bodies and $X$ stands for the
central body of the XCRS. Here we deliberately
replaced all Earth-related quantities, denoted with subscript '$E$' in
the IAU 2000 framework, by the quantities corresponding to the
arbitrary central body denoted by subscript '$X$'.  All the details of
various terms in these transformations can be found e.g. in
\citet{2003AJ....126.2687S}.

The validity region of the local coordinate systems is mathematically
only limited by the BCRS acceleration $\ve{a}_X$ of the central body, so that 
$|\ve{X}|<c^2/|\ve{a}_X|$. This easily covers the whole Solar system
even for Mercury with its acceleration $\sim0.063$\,m\,s$^{-2}$.
However, from the physical point of view, the local GCRS-like reference
systems are useful and more convenient than the BCRS only within the
immediate vicinity of the corresponding central body. The main role of the local
reference systems is to adequately describe the dynamics of the central
bodies (e.g. their rotational motions), the multipole structure of
their gravitational fields, the geodesy of their surface as well as
the motions of spacecrafts orbiting those bodies. For the GCRS, its
use should be limited to the region within, say, the geostationary
orbit (i.e. within 7-10 radii of the Earth from the geocenter). The
LCRS remains clearly useful within $\sim10$ lunar radii from the center of
the Moon. Therefore, the use of the LCRS for the whole 'cislunar'
space (understood as the region of space within the lunar orbit)
cannot be recommended. Outside the immediate vicinity of the Moon and
the Earth, the BCRS and its coordinate time \TCB\ should be used and are
already used to describe e.g. the geocentric orbit of the Moon in all modern
solar system ephemerides.

\section{Two kinds of time scales in general relativity}

One should clearly distinguish between two kinds of time scales in the relativistic context:

\smallskip

1) {\sl Coordinate time} is one of the four coordinates of a
four-dimensional relativistic reference system, e.g., BCRS, GCRS, LCRS
etc. The coordinate time of a given reference system
is defined for any four-dimensional event in the
region of space-time covered by that reference system.  An event is
some physical phenomenon localized both in time and in space.
Different coordinate times for the same four-dimensional event
can be transformed one into another using,
generally speaking, some four-dimensional transformations.  In the IAU
framework the transformation between the coordinate time $t=\TCB$ of
the global BCRS and the coordinate times $T$ (e.g. \TCG\ or \TCL\ for
the GCRS or LCRS, respectively) are given by Eq.~(\ref{T-t-IAU})
above. One can immediately see that the transformation depends on the
spatial position $x^i$ of the event for which the transformation is
computed. 

\smallskip

2) {\sl Proper time} $\tau_{\rm obs}$ of an observer is the reading of
an ideal clock located and moving together with the observer. The
proper time is meaningful only for a specific observer and can be
related to the coordinate time $t$ of any reference system that covers
the relevant part of the trajectory of the observer ('$t$' stands here
for any coordinate time -- both \TCB\ and any of the \TCG-like
coordinate times). This relation reads:
  \begin{eqnarray}
    {d \tau_{\rm obs}(t)\over dt}=&\biggl(&-g_{00}(t,\ve{x}_{\rm obs}(t))
\nonumber \\
  &&-{2\over c}g_{0i}(t,\ve{x}_{\rm obs}(t))\,\dot{x}^i_{\rm obs}(t)
-{1\over c^2}g_{ij}(t,\ve{x}_{\rm obs}(t))\,\dot{x}^i_{\rm obs}(t)\,\dot{x}^j_{\rm obs}(t)\biggr)^{1/2}\,,
\label{eq:dtaudt}
  \end{eqnarray}
  \noindent
  where $g_{\alpha\beta}(t,\ve{x})$ are the components of the metric
  tensor in the corresponding reference system with coordinates
  $(t,x^i)$ (this can be both the BCRS or any of the local reference
  systems XCRS), $\ve{x}_{\rm obs}(t)$ in the trajectory of the
  observer in this reference system. Eq.~(\ref{eq:dtaudt}) only
  defines the derivative of $\tau_{\rm obs}$ with respect to the
  coordinate time, so that an initial condition
  \begin{equation}
    \label{t0-tau0}
  \tau_{\rm obs}(t_0)=\tau_0
  \end{equation}
  \noindent
  is required to evaluate $\tau_{\rm obs}$ as function of $t$. Here $t_0$ and $\tau_0$ are some constants selected from some operational needs or
  considerations of convenience.

\smallskip

Coordinate time scales represent the only way to define simultaneity
and clock synchronization in General Relativity (even in the post-Newtonian
approximation) for the complex case of realistic Solar system or for
the vicinity of the Earth.  The so-called coordinate simultaneity and
synchronization were introduced by \citet{1979RaSc...14..649A} and
used to formulate practical synchronization algorithms by many authors
\citep{1992CeMDA..53...81K,1994A&A...286..971P,1995A&A...304..653W,2001A&A...370..320B}.
The basic idea is to use special sort of observation (one-way or
two-way synchronization etc.) to compute the initial conditions
(\ref{t0-tau0}) in such a way that each clock with its proper time
realizes a single coordinate time. This procedure can be performed to
implement \TCB\ or any local coordinate time scale \TCX\ with an ensemble of
real clocks moving along arbitrary trajectories.

Another interesting point is the relation between the proper time of
an observer located at the origin of the local reference system and
the corresponding coordinate time. Since the observer is located at
the origin, its velocity in this reference system vanishes and
only one term survives in Eq.~(\ref{eq:dtaudt}) to give
  \begin{equation}
    {d \tau^{\rm origin}_{\rm obs}(T)\over dT}=\left(-G_{00}(T,\ve{X})\right)^{1/2}\Biggr|_{\ve{X}\equiv0}=
    1-{1\over c^2}\,W_X(T,{\bf 0})-{1\over 2c^2}\,W_X^2(T,{\bf 0})+O(c^{-5})\,,
\label{eq:dtaudt-origin}
  \end{equation}
\noindent
Here $G_{00}$ is the corresponding 'time-time' component of the metric tensor in
the considered local reference system and $W_X(T,{\bf 0})$ is the
gravitational potential of the central body at its center of mass
where $\ve{X}={\bf 0}$.  Obviously, the most important contribution here
comes from the Newtonian gravitational potential, which is,
contrarily a frequent misconception, by no means zero at the center of
mass of a body. For the real major bodies of the Solar system like the
Earth and the Moon, we don't know the values of the gravitational
potential at the respective center of mass. However, for a model of homogeneous sphere
one can show that the Newtonian gravitational potential is
time-independent, depends only on the distance
$r=|\ve{X}|$ from the center of mass, and reads:
\begin{equation}
  W_X=\left[\phantom{{\LARGE\Bigg[}}\hspace*{-2mm} 
{\begin{array}{ll}
  \displaystyle{GM\over r}, & r\ge L  \\[8pt]
  \displaystyle{GM\over 2L^3}\,\left(3L^2-r^2\right), & r<L
\end{array}}  
    \right.
\end{equation}  
\noindent
where $L$ is the radius of the sphere, $M$ its mass, $G$ is the
Newtonian gravitational constant, and the relativistic terms are
neglected.  So, at the center of mass the gravitational potential is
$W_X(T,{\bf 0})={3GM\over 2L}$ which a factor of 3/2 larger than the
potential on the surface $GM/L$.  Therefore, the proper time
$\tau^{\rm origin}_{\rm obs}$ don't coincide with the respective
coordinate time at the origin of the coordinates. This is also true
for the BCRS since the gravitational potential at the solar system
barycenter is also by no means zero.  One should anyway bear in mind
that this consideration only involves one particular observer while
coordinate times are defined for any events in the whole Solar system
-- this makes coordinate times anyway distinct from proper times of any
observers.

\section{It is unreasonable to scale \TCL}

Mostly for historical reasons and because the vast majority of
high-accuracy and high-precision (highly stable) clocks are located on
the surface of the Earth, scaled coordinate times $\TT=(1-L_G)\,\TCG$
and $\TDB=(1-L_B)\TCB+\TDB_0$ are defined in the IAU Resolutions and used in practice. Here
$L_G$, $L_B$ and $\TDB_0$ are certain constants. The ideas
behind those scaled time scales are that (a) the rate of \TT\ agrees
with that of the proper time for an observer $\tau_{\rm obs}$ on the
rotating geoid with very specific defining constant $W_0=c^2\,L_G$ \citep[see][]{2003AJ....126.2687S}
and (b) the scaled \TDB\ remains as
close as possible to \TT\ evaluated at the geocenter: the linear drift is approximately removed by the scaling and
the main part of
the remaining difference is quasi-periodic with an amplitude of about 2
milliseconds. From the practical point of view, the relatively small
differences between \TDB\ and \TT\ can be ignored for many
applications localized in the vicinity of the Earth. These scaled times
improve the security e.g. of spacecraft operations since an
accidental use of e.g. \TT\ instead of \TDB\ is unlikely to lead to some
catastrophic results.

However, each scaling of the coordinate time implies scaling of the
spatial coordinates, the mass parameters, and some other parameters
like the Maxwell-type multipole moments. Already now we must
distinguish between three values of all mass parameters: the physical
value (\TCG- and \TCB-compatible), \TT-compatible value and
\TDB-compatible value (see Table 1 of \citet{2011CeMDA.110..293L}). A
detailed discussion of the scalings of coordinate times and their
consequences can be found in \citet{2008A&A...478..951K} (see also
\citet{2010IAUS..261...79K} where it is argued that the scalings don't
change the units of measurements: according of the IAU resolutions,
the SI second and SI meters are used for all coordinate and observed quantities).

Already with the introduction of the third coordinate time \TCL\ (as a
part of LCRS) it is no longer possible to eliminate linear drifts
between \TT, \TDB\ and a scaled version of \TCL: linear drifts will
remain present whichever scaling constants would be chosen. Given also
the implication of the additional scaling of spatial coordinates and
astronomical constants, it turns out to be unnecessary, unreasonable
and even risky to scale also the coordinate times of other GCRS-like
coordinate systems like \TCL\ or its analogy for Mars or Mercury.

\section{Mathematical scheme for the coordinate time transformations}

Here we formulate practical recipes to work with the transformations
between \TCB\ and the coordinate times \TCX\ of the local GCRS-like
reference systems for any bodies of the Solar system. The approach
described below was used as a part of the \gaia\ timing system
\citep{GAIA-CA-TN-LO-SK-012}.

The transformations between the coordinate time $t=\TCB$ of the BCRS
and the local coordinate time $T=\TCX$ for any body $X$ are given by
Eq.~(\ref{T-t-IAU}) above. One sees that there are two kinds of terms
here: (a) the terms depending on the BCRS position $\ve{x}$ of the
event for which the transformation is computed and (b) the terms that
don't depend on the position. The first kind of terms can be estimated
depending on the maximal distance between the center of mass of the
central body and the event where the transformation is performed. For
\TCG\ such estimates are given in Table~I of
\citet{GAIA-CA-TN-LO-SK-012}. For Mercury these estimates are given by
\citet{FallettaRodridezKlioner2025}. For the Moon, the estimates are
given in Table~\ref{tab:position-terms-Moon}. It can be seen that in the
foreseeable future most probably one only needs to consider for
\TCL\ the post-Newtonian terms $-c^{-2}\,v_L^ir_L^i$ even if the whole
cislunar space is considered (what is not recommended -- see above).

%
%

%
%

\begin{table}[h!]
 \centering
 \caption{Estimations of the location-dependent terms in the \TCL--\TCB\ transformation.}
\label{tab:position-terms-Moon}
 {\tablefont\begin{tabular}{@{\extracolsep{\fill}}lrr}
   \midrule
   & \hfill $|\ve{r}_L|<R_L\approx 1.74\times10^6$\,m\hfill &$|\ve{r}_L|<2R^{\rm max}_{EL}\approx 8.11\times10^8$\,m\hfill\\
   \midrule
   $-c^{-2}\,v_L^ir_L^i$                & $6.1\times 10^{-7}$\,s  & $2.9\times 10^{-4}$\,s \\
   $+c^{-4}\,B^i(t)r_L^i$               & $2.2\times 10^{-14}$\,s & $1.0\times 10^{-11}$\,s \\
   $+c^{-4}\,B^{ij}(t)r_L^i r_L^j$       & $1.1\times 10^{-19}$\,s & $2.3\times 10^{-14}$\,s \\
   $+c^{-4}\,C(t,\ve{x})$              & $7.2\times 10^{-25}$\,s & $7.3\times 10^{-17}$\,s \\ 
    \midrule
    \end{tabular}}
 \tabnote{\textit{Notes}: Here $\ve{r}_L=\ve{x}-\ve{x}_L$, $\ve{x}_L$
   is the BCRS position of the Moon, $\ve{v}_L$ is the BCRS velocity
   of the Moon, $R_L$ is the radius of the Moon, and $R^{\rm
     max}_{EL}$ is the maximal distance between the Earth and the
   Moon. The formulas for $B^i$, $B^{ij}$, and $C(t,\ve{x})$ are given as Eqs.~(\ref{eq:Bi})--(\ref{eq:C})}.
\end{table}

The position-dependent terms vanish at the origin of the local reference
system, the BCRS coordinate of which are given by $\ve{x}_X$ for body
$X$.  At the origin of the local reference system (called the
'geocenter' or the 'center of mass of the Moon' etc.) only two terms
in Eq.~(\ref{T-t-IAU}) survive: $-c^{-2}\,A(t)$ and $+c^{-4}\,B(t)$
defined by Eqs.~(\ref{At})--(\ref{Bt}). Therefore, one has
\begin{eqnarray}
\label{TCG-TCB}
{dT_{\rm c}\over dt}&=&1+F(t)\,,\\
\label{F-def}
F(t)&=&-{1\over c^2}\,\dot{A}(t)+{1\over c^4}\,\dot{B}(t)\,.
\end{eqnarray}
where index '${\rm c}$' stresses that the coordinate time $T$ is
evaluated at the center of mass of the central body (that is, at the
origin of the corresponding local reference system). The solution of Eq.~(\ref{TCG-TCB})
$T_{\rm c}(t)$ and its inversion are called time ephemerides for the central body $X$.
Since the distances between the major solar system bodies are sufficiently large, one can
neglect the non-sphericity of the gravitational fields of the bodies when computing $F(t)$.
Thus, assuming a monopole gravitational field for each body one has
\begin{eqnarray}
\label{dot-A}
\dot{A}(t)&=&{1\over 2}\,v_X^2+\sum_{A\neq X}{GM_A\over r_{XA}},
\\
\label{dot-B}
\dot{B}(t)&=&-{1\over 8}\,v_X^4
+{1\over 2}\,
{\left(\,\sum_{A\neq X} {GM_A\over r_{XA}}\right)}^2
+
\sum_{A\neq X} \left({GM_A\over r_{XA}} \sum_{B\neq A} {GM_B\over r_{AB}}\right)
\nonumber\\
&&
+\sum_{A\neq X} {GM_A\over r_{XA}}
\Biggl(4\,
v_A^i v_X^i
-{3\over 2}\,
v_X^2
-2\,
v_A^2
+{1\over 2} a_A^i r_{XA}^i
+{1\over 2} (v_A^i r_{XA}^i/r_{XA})^2 \Biggr),
\end{eqnarray}
\noindent
where capital Latin subscripts $A$, $B$ and $C$ enumerate massive
bodies, $X$ corresponds to the central body, $M_A$ is the mass of body
A, $\ve{r}_{XA}=\ve{x}_X-\ve{x}_A$, $r_{XA}=|\ve{r}_{XA}|$,
$\ve{v}_A={\dot {\ve{x}}}_A$, $\ve{a}_A={\dot {\ve{v}}}_A$, a dot
signifies time derivative with respect to $t=\TCB$, and $\ve{x}_A$ is
the BCRS position of body A.
%

In order to compute both $T_{\rm c}(t)$ and its inverse function two functions $\Delta\,t(t)$ and $\Delta T(T)$
are defined (here and below the subscript '${\rm c}$' is omitted):
\begin{eqnarray}
\label{Delta-t-def}
T&=&t+\Delta\,t(t),
\\
\label{Delta-T-def}
t&=&T-\Delta T(T).
\end{eqnarray}
\noindent
with two ordinary
differential equations for $\Delta\,t(t)$ and $\Delta T(T)$:
\begin{eqnarray}
\label{Delta-t}
{d\Delta\,t\over dt}&=&F(t),
\\
\label{Delta-T}
{d\Delta T\over dT}&=&{F(T-\Delta T(T))\over 1+F(T-\Delta T(T))}.
\end{eqnarray}
\noindent
These relations are exact for some given $F(t)$. Eq.(\ref{F-def}) for
$F$ is obviously approximate.  Initial conditions for these two
differential equations are given by the IAU definitions of \TCB, \TCG,
and \TCL\ and assumed for all other \TCX: $\TCB=\TCX=32.184$ s on
1977, January 1, 0$^h$ 0$^m$ 0$^s$ \TAI\ at the geocenter. In terms of
Julian Dates $JD_{\,\TCB}$ and $JD_{\,\TCX}$ in \TCB\ and \TCX, respectively
one has:
\begin{eqnarray}
\label{Delta-t-zero}
\Delta\,t(JD_{\,\TCB} = 2443144.5003725)&=&0,
\\
\label{Delta-T-zero}
\Delta T(JD_{\,\TCX}= 2443144.5003725)&=&0.
\end{eqnarray}
\noindent
Any reasonable integrator for ordinary differential equations can be
used to integrate the differential equations
(\ref{Delta-t})--(\ref{Delta-T}) with given initial conditions
(\ref{Delta-t-zero})--(\ref{Delta-T-zero}). The accuracy of numerical
integrations can be automatically checked, e.g., by integrating forth
and back and comparing the results. The consistency of two independent
integrations (one for $\Delta\,t(t)$ and another one for $\Delta
T(T)$) can be cross-checked using identities $\Delta\,t(t)\equiv
\Delta T(t+\Delta\,t(t))$ and $\Delta T(T)\equiv \Delta\,t(T-\Delta
T(T))$.

We note that the same mathematical approach can be used to compute the
transformation between the proper time of some observer $\tau_{\rm
  obs}$ and a coordinate time $t=\TCB$ along the trajectory of that
observer $\ve{x}_{\rm obs}(t)$.  To this end, one considers the
observer as the 'center body' with negligible mass $M_X=0$. This
latter assumption slightly simplifies Eq.~(\ref{dot-B}) that defines
the derivatives of $B(t)$.  This approach was used in the \gaia\ data
processing \citep{GAIA-CA-TN-LO-SK-012}.  The time ephemeris of the
\gaia\ spacecraft is part of the algorithm for the calibration of the
\gaia\ on-board Rb clock
\citep{2015jsrs.conf...55K,2017SSRv..212.1423K}. The \gaia\ clock
calibration will be published in the \gaia\ Data Release 4.

\section{Numerical time ephemerides for the major Solar system bodies}

Using the mathematical model described above, the functions
$\Delta\,t(T)$ and $\Delta T(t)$ were numerically computed for all
major bodies of the Solar system included in the export version of the
solar system ephemeris INPOP19a \citep{2019NSTIM.109.....F}: the eight
major planets, Pluto, the Moon and the Sun. The \TCB-compatible
version of the INPOP19a ephemeris valid for 200~years centered on
J2000 was used to compute the dynamical model for $F(t)$. The INPOP19a
was downloaded in the form of ASCII files from the web page of the INPOP team.
The resulting functions $\Delta\,t(t)$ and $\Delta T(T)$ were
represented as a set of Chebyshev polynomials in the standard form that
is also used e.g. for the INPOP ephemerides themselves. The resulting time
ephemerides together with some visualizations can be found in
\hbox{\small\url{https://gaia.geo.tu-dresden.de/TimeEphemerides/}} (with a
more persistent copy at
\hbox{\small\url{https://doi.org/10.5281/zenodo.17691611}}).

The general integrator ODEX was used to integrate
Eqs.~(\ref{Delta-t})--(\ref{Delta-T}) with initial conditions
(\ref{Delta-t-zero})--(\ref{Delta-T-zero}). The numerical errors of
the numerical integration as well as Chebyshev approximation were kept
well below 10 picoseconds. As the model for $F(t)$
Eqs.~(\ref{dot-A})--(\ref{dot-B}) with all bodies included in INPOP19a
were used. Also the mass parameters from INPOP19a were used.

This model for $F(t)$ is limited to the export version of INPOP19a and also
doesn't include the gravitational potential of asteroids or the
multipole structure of the gravitational field. Those additional
forces were used internally to create INPOP19a
\citep{2019NSTIM.109.....F}.  The ephemeris providers have full access
to that part of the dynamical model and should in principle be able to
produce more accurate time ephemerides in the same way it is done
already for the \TCG--\TCB\ and \TT--\TDB\ relations at the
geocenter. From this point of view, the time ephemerides presented
here can be considered as a demonstrators.  We note also that the
parameters of the Chebyshev representations were not fully optimized
and most probably the order of the polynomials for each granule can be
further lowered.

Table~\ref{table-linear-periodic} shows some characteristics of the
time ephemerides for various solar system bodies. Functions $\Delta
t(T)$ and $\Delta T(t)$ can be approximately represented as a sum of a
linear drift and a quasi-period function, the magnitude of which are
given in the table.  No such split is required or even useful for practical
work. It should be stressed that this split is only
approximate. Indeed, the results of a fit of a linear drift depends on
the particular time period used in the fit, on whether some periodic
terms are fitted simultaneously with the linear drift, etc. For this
reason, the figures are given with a low precision in
Table~\ref{table-linear-periodic}.

In order to estimate the realistic uncertainties of the time
ephemerides presented here (the uncertainties that come from the fact
that the model for $F(t)$ is truncated), the time ephemeris $\Delta
T(t)$ for the Earth computed here as compared to the same time
ephemeris delivered as a part of INPOP19a. It was found that the
difference is a linear drift of about $6.2\times10^{-18}$ (this
corresponds to a time difference of 19.5 nanoseconds over 100 years)
and quasi-periodic terms with an amplitude of 0.15 nanoseconds.

\begin{table}[h!]
 \centering
 \caption{Approximate magnitudes of the linear drifts and quasi-periodic terms in the time ephemerides}
 \label{table-linear-periodic}
 {\tablefont
    \begin{tabular}{@{\extracolsep{\fill}}lll}
    \midrule
    Body& linear drift& maximal amplitude of\\
    & [$\times10^{-8}$] & quasi-periodic terms [ms]\\
    \midrule
    Sun & 0.000222 & 0.0076 \\
    Mercury & 3.825 & 12.7 \\
    Venus & 2.047 & 0.60 \\
    Earth & 1.481 & 1.69 \\
    Moon & 1.483 & 1.80 \\
    Mars & 0.972 & 11.5 \\
    Jupiter & 0.285 & 11 \\
    Saturn  & 0.155 & 18 \\
    Uranus  & 0.077 & 22 \\
    Neptune & 0.049 & 5.5 \\
    Pluto   & 0.043 & $\sim$100\\
    \midrule
    \end{tabular}
  }
\tabnote{\textit{Notes}: The values for the Earth are given for completeness. The corresponding linear drift is {\it approximately} equal to the constant $L_C$ \citep[see e.g.][]{2011CeMDA.110..293L}.}
\end{table}

\section{Concluding remarks}

This paper overviews the properties and the intended use of the
relativistic coordinate times as they are defined in the IAU 2000
framework. It is argued that the GCRS-like relativistic local
reference systems as well as the \TCG-like local coordinates times are
already defined for each body of the Solar system. Officially defined
25 years ago, the IAU 2000 framework still represents the best possible choice
for the definition of the relativistic reference systems that should
be used in the algorithms of relativistic modelling of all kinds of
high-accuracy observational data in astronomy.

We argue that further scaling of these coordinate times (similar to
the scaling between e.g. \TT\ and \TCG) is unreasonable and risky from
the operational point of view: each such scaling implies a set of
scaled astronomical constants and scaled spatial coordinates that
differ from what is being used now. Moreover, no additional scaling of
\TCL\ can remove linear drifts between all relevant coordinate time
scales.

We also argue that the GCRS-like local reference system, e.g. LCRS,
should only be used in the immediate vicinity of the central body
(say, up to 10 radii of the corresponding body). Outside this region
the BCRS should be used.  In particular, the BCRS should be used in
the so-called cislunar space (the space region bounded by the lunar
orbit). This also corresponds to the current practice of using the
BCRS coordinates to describe the geocentric orbit of the Moon in all
modern solar system ephemerides.

Finally, we present a practical algorithm to compute the time
ephemerides, that is, the numerical transformations of the local \TCG-like
coordinate times and \TCB\ evaluated at the center of mass of the
corresponding central body. These transformations were numerically
computed using INPOP19a as the underlying theory and made available
as standard sets of Chebyshev polynomials.

\end{document}